\documentclass[conference]{IEEEtran}
\usepackage[T5]{fontenc}

\usepackage{cite}
\usepackage{amsmath,amssymb,amsfonts}
\usepackage{algorithmic}
\usepackage{graphicx}
\usepackage{textcomp}
\usepackage{xcolor}
\usepackage{todonotes}
\usepackage{xspace}
\usepackage{hyperref}
\usepackage{cleveref}

\def\BibTeX{{\rm B\kern-.05em{\sc i\kern-.025em b}\kern-.08em
    T\kern-.1667em\lower.7ex\hbox{E}\kern-.125emX}}

\usepackage{orcidlink}

\usepackage{comment}

\usepackage{eiffel}

\newcommand{\e}[1]{\mbox{\lstinline[basicstyle=\normalsize]|#1|}}

\begin{document}

\title{\toolname: a standard language, database schema and repository for research on bugs and automatic program repair}

\newcommand{\toolnamefull}{Bugs Are Definetely A Serious Stuff\xspace}
\newcommand{\toolname}{Bugfix\xspace}
\newcommand{\sitelink}{\url{https://icbf813.github.io/}\xspace}

\author{\IEEEauthorblockN{Victoria Kananchuk\,\orcidlink{0009-0000-3008-6245}}
\IEEEauthorblockA{\textit{Constructor Institute of Technology} \\
Schaffhausen, Switzerland \\
Viktoryia.Kananchuk@constructor.org}
\and
\IEEEauthorblockN{Ilgiz Mustafin\,\orcidlink{0009-0007-0476-5966}}
\IEEEauthorblockA{\textit{Constructor Institute of Technology} \\
Schaffhausen, Switzerland \\
Ilgiz.Mustafin@constructor.org}
\and
\IEEEauthorblockN{Bertrand Meyer\,\orcidlink{0000-0002-5985-7434}}
\IEEEauthorblockA{\textit{Constructor Institute of Technology}\\
Schaffhausen, Switzerland \\
Bertrand.Meyer@inf.ethz.ch}
}

\maketitle

\begin{abstract}\footnote{An early step towards this article was a short contribution \cite{aprlisbon} to the 2024 Automatic Program Repair at the International Conference on Software Engineering (ICSE 24). The present text reuses a few elements of introduction and motivation but is otherwise thoroughly reworked and extended.}
Automatic Program Repair (APR) is a brilliant idea: when detecting a bug, also provide suggestions for correcting the program. Over the past decades, researchers have come up with promising techniques for making APR a mainstay of software development. Progress towards that goal is, however, hindered by the absence of a common frame of reference to support and compare the multiplicity of ideas, methods, tools, programming languages and programming environments for carrying out automatic program repair.

\toolname, described in this article, is an effort at providing such a framework: a standardized set of notations, tools and interfaces, as well as a database of bugs and fixes, for use by the APR research community to try out its ideas and compare the results on an objective basis.

The most directly visible component of the \toolname effort is the \toolname language, a human-readable formalism making it possible to describe elements of the following kinds: a \textit{bug} (described abstractly, for example the permutation of two arguments in a call); a bug \textit{example} (an actual occurrence of a bug, in a specific code written in a specific programming language, and usually recorded in some repository); a \textit{fix} (a particular correction of a bug, obtained for example by  reversing the misplaced arguments); an \textit{application} (an entity that demonstrates how a actual code example matches with a fix); a \textit{construct} (the abstract description of a programming mechanism, for example a ``while'' loop, independently of its realization in a programming language; and a \textit{language}  (a description of how a particular programming language includes certain constructs and provides specific concrete syntax for each of them --- for example Java includes loop, assignment etc. and has a defined format for each of them).


The language is only one way to access this information. A JSON-based
\textit{program interface} (API) provides it in a form accessible to tools, which may record and
access such information into databases. \toolname includes such a database
(\textit{repository}), containing a considerable amount of bugs, examples and fixes.

We hope that this foundational work will help the APR research community to advance the field by providing a common yardstick and repository.

\end{abstract}


\section{Introduction} \label{introduction}


Techniques of software verification, from tests to static analysis and proofs, help identify bugs; in recent years the idea has emerged that while spotting a bug is good, devising a possible \textit{correction} for the bug is better. This approach has come to be known as ``Automatic Program Repair'' or APR. 

In practice, Automatic Program Repair techniques usually do not actually perform changes to the program but \textit{suggest} one or more possible corrections to the programmer, who remains in charge of choosing and applying one of them. The basic toolset that programmers use (Interactive Development Environment, or IDE) should include tools that detect bugs and, for example, pop up a message that signals the bug and proposes one or more valid corrections. The present article does not examine these (important) practical matters but concentrates on the underlying techniques making such a scenario possible. 

Interest in APR techniques has developed considerably in the past two decades, starting with early work such as \cite{liu2018lsrepair, nguyen2013semfix, 7463060, 6776507, 7203042, 10.1145/3180155.3180250}. There is now a considerable literature; good surveys can be found in \cite{monperrus2018automatic, 
monperrus2018living, gazzola2018automatic}. 

One of the major obstacles to  widespread deployment and use of  APR tools is the heterogeneity of the field: any solution must deal with the wide diversity of APR approaches, of IDEs, of programming languages, and also of bugs. These factors force every APR project to invest a major part of its effort in recreating a basic bug and fix description framework; such spurious repetition of work considerably impair progress towards making Automatic Program Repair a standard and effective part of the software developer’s daily experience.

The \toolname project, described in the present article, addresses this obstacle by  establishing a joint framework, \toolname, that all APR efforts can use. The starting point for \toolname was a short position paper \cite{aprlisbon} at the at the APR workshop of ICSE. The present proposal makes a number of different design choices and its description below provides enough detail to allow testing and APR projects to start using it.

It includes mechanisms for specifying, in a standard way, elements of six different kinds. The first four pertain to bugs and their fixes:

\begin{itemize}
    \item A \e{bug} element, describing a bug pattern, such as using a strict
      comparison
      (\e{i < n}) instead of a nonstrict-one (\e{i <= n}) (section \ref{bugs}).
    \item A \e{fix} pattern for a bug pattern, such as replacing the nonstrict operator by the strict version (section \ref{fixes}).
    \item An \e{example} of a fix in a particular program which shows the before-
      and after- states of the code (section \ref{examples}).
    \item An \e{application} of a fix to an example which shows how exactly the
      bug pattern matches the before-state of the code and how the fix pattern
      transforms the code to the after-state (section \ref{applications}).
\end{itemize}

\noindent Elements of these kinds are described independently of a particular
programming language (although some of them may be meaningful in some
programming languages only, as in the case of bug in a ``repeat... until...''
loop, which can only occur in a language such as Pascal supporting a loop of
that kind). Actual bugs occur in actual programming languages; the remaining two
kinds of element enable \toolname to achieve full generality by specifying the constructs in which bugs can occur and fixes can be applied:

\begin{itemize}
    \item A \e{construct} element (\ref{constructs}) describes a particular programming language construct,  such as the ``repeat... until...'' loop. Although such a description is guided by the availability of a construct in existing programming languages, it is still language-independent: it provides a structural description of the construct (an abstract syntax), by listing its components, such as an exit condition and a loop body in this case, without providing any particular concrete syntax.
    \item A \e{language} specification (\ref{language}) indicates how a programming language uses particular constructs. It specifies the list of constructs in the language (at least those for which bugs have been identified) and, for each of these constructs, what concrete syntax it uses. For example Java includes loops, assignments, conditionals etc. and provides a specific way of expressing each of them, such as ``\e{target = source;}'' for an assignment, where \e{source} and \e{target} are components of the corresponding (language-independent) \e{construct} specification. 
\end{itemize}

These six kinds of element provide the core mechanisms of \toolname. They are available in various ways. In particular:

\begin{itemize}
    \item The description of the mechanism (in sections \ref{bugfix1} and
      \ref{bugfix2}) uses the \toolname \textbf{language}, a clear human-readable notation best suited for learning the concepts and discussing bugs, examples, fixes, constructs and languages.

    \item These mechanisms must also be usable through programs, or through queries on bug and fix databases. For that purpose, a \textbf{program interface (API)} is available, using the JSON conventions (section \ref{API}). The functionality it provides is the same as the functionality of the language.

    \item To illustrate the approach and provide a directly available resource,
      we have collected a large number of bugs, examples and fixes in a publicly
      available \textbf{repository} (section \ref{API}).
\end{itemize}

\section{The \toolname language} \label{bugfix}

At the core of \toolname lies a language for describing identified patterns for both bugs and their fixes. Two observations:
\begin{itemize}
    \item The \toolname effort focuses on bugs that manifest themselves through code patterns that are wrong and should be replaced (fixed) by adapting the patterns. For example, code that uses \e{f (a, b)} when it should use \e{f (b, a)}. We do not at this point consider deeper or more elaborate bugs, such as design bugs, as analyzed for example in \cite{catolino2019not}.
    
    \item To illustrate the \toolname language, we use a concrete syntax. For clarity the syntax is keyword-oriented. Such concrete syntax details are not essential; as noted, the same functionality is available through an API (section \ref{API}. Other versions of the syntax are possible.
    
\end{itemize}

Section \ref{bugfix1} presents bug-related elements: \e{bug} (\ref{bugs}),
\e{fix} (\ref{fixes}), \e{application} (\ref{applications}) and \e{example} (\ref{examples}). Section \ref{bugfix2} presents language-related elements: \e{construct} (\ref{components}) and \e{language} (\ref{language}).

The \toolname language description is informal, through examples; a full language reference will be available on the \toolname website included as part of the Artifacts.

\section{Bugs, fixes, applications and examples} \label{bugfix1}
Part of the \toolname language is dedicated to specifying bug patterns, examples of
such patterns, and proposed fixes.

These descriptions are language-independent and based on the idea of a
universal Abstract Syntax Tree (AST) which consists of constructs
(specified per the rules of \ref{constructs} below).

\subsection {Specifying bug patterns} \label{bugs}
A bug specification presents a certain program pattern, of which some instances have been identified as buggy (although not all instances necessarily are).

Such a specification describes the general scheme for a bug,
not a particular occurrence.
Defining specific occurrences is the role of \textit{application} specifications (\ref{applications}), each of which refers to a given \e{bug} specification. (As seen in \ref{together} below, it is also possible, for convenience, to specify an example together with the corresponding bug.)

The following is an example of bug specification:

\begin{lstlisting}
bug WRONG_ARGUMENT_IN_CALL_1
    -- In a call with an actual argument list
    -- starting with `pre' and
    -- ending with `post', 
    -- the actual argument between these parts
    -- may not be the right one.
    -- So in `f (a, b, m, u, v)'
    -- `m'is not right.
    -- Here `initial' is `a, b',
    -- `final' is `u, v',
    -- `wrong_arg' is `m'.
    -- Note that `pre' and `post'
    -- may each be empty.
where
  (ARGUMENT_LIST
    (_)* @pre
    (_) @wrong_arg
    (_)* @post)
)
end
\end{lstlisting}

The specification defines a bug pattern called
\e{WRONG_ARGUMENT_IN_CALL_1} which matches any \e{ARGUMENT_LIST} node
which has at least one argument (zero or more \e{@pre} arguments,
one \e{@wrong_arg} argument and zero or more \e{@post} arguments).

The syntax used for the pattern is based on S-expressions (sexps)
and is similar to the query syntax of the tree-sitter \cite{tree-sitter}.
Tree-sitter is discussed more in \cref{related_work}.

Each sexp starts with a name of a construct or a kind or which must
match at this position. The wildcard symbol ``\e{\_}'' can be used
for matching any node.

The following elements of sexp define the children of the node. Each child
definition starts with an optional construct feature name like \e{right:},
followed by a required sexp specifying the pattern, followed by
an optional quantifier
(such as ``\e{*}'' for 0 or more occurences, ``\e{+}'' for 1 or more, ``\e{?}''
for 0 or 1), followed by an optional capture name like \e{@wrong}.

Nodes which have a specified capture name can be used in \textbf{fixes}.
The top-level node always has an implicit capture name \e{@bug}.

\subsection {Specifying fixes} \label{fixes}
A fix specification presents a proposed replacement for a bug pattern defined by a
\e{bug} specification per \ref{bugs}.
Since a fix specification always describes a fix for a particular bug, it will refer explicitly to that bug in the \e{for} clause. (An alternative is to declare the fix with the bug, as explained in the next subsection.)

The fix specification uses the \e{parameter} clause to list the required constructs
which are needed to fill the blanks in the fix;
and the \e{then} clause to specify the replacement subtree
for the whole pattern captured by the \e{bug}.
For example:

\begin{lstlisting}
fix CORRECT_ARGUMENT_IN_CALL_2
    -- In the call,
    -- replace the `@wrong_arg' argument
    -- by `@correct_arg'.
bug_id
    WRONG_ARGUMENT_IN_CALL_2
parameter
  correct_arg: EXPRESSION
then
  (ARGUMENT_LIST
    @initial
    @correct_arg
    @final)
end
\end{lstlisting}

The \e{feature} clause specifies that a node of type \e{EXPRESSION} named
\e{@correct_arg} is used in the \e{then} clause.

The \e{then} clause defines the replacement subtree for the subtree captured by
the \e{bug} pattern. The syntax used in the \e{then} part is also based on
sexps: the first element is a name of the construct followed by the child
definitions. Child definition starts with an optional feature name like
\e{right:} and is followed by a sexp, a bug capture name or a fix feature
name; or a string value for the leaf nodes.

An alternative syntax for might be used for splicing (splatting in Ruby or
spreading in JavaScript) captured nodes and fix features using the splice
symbol ``\e{*}''. A sexp starting with the splice symbol followed by a name
is equivalent to a sexp with the same contents as the node represented by the
name. Child nodes can be overriden by simply specifying them in the sexp.

Splice symbol can appear after the construct name as in \e{(X *@y)} and
is equivalent to a sexp with the construct name set as \e{X} and children
as in the node represented by \e{@y}.

\subsection {Specifying applications} \label{applications}
An \e{application} specification presents how a real bug occurence in an
\textbf{example} is captured by a \textbf{bug} pattern
and how the \textbf{fix} pattern produced the fixed version of the
\textbf{example}.

The \e{tree} clause shows how the real code is represented by the
construct-tree and which nodes are captured with capture names.

The \e{parameter} clause specifies the values for \e{fix} features which
lead to the fix matching the \e{example}.

For example:
\begin{lstlisting}
application CORRECT_ARGUMENT_IN_CALL_1_CLOSURE_14 
example
  CLOSURE_14
fix
  CORRECT_ARGUMENT_IN_CALL_1
tree
  argument_list [766, 28] - [766, 66]
    identifier [766, 29] - [766, 37] @pre
    method_invocation[766, 39] - [766, 52] @wrong_arg
      object: identifier [766, 39] - [766, 45]
      method: identifier [766, 46] - [766, 52]
    identifier [766, 54] - [766, 65] @post
parameter
  @correct_arg = (field_access
    object: (identifier "Branch")
    field: (identifier "ON_EX"))
end
\end{lstlisting}

The specification presents an application of the fix
\e{CORRECT_ARGUMENT_IN_CALL_1} which matches the real code \textbf{example}
\e{CLOSURE_14}. The bug pattern matches the \e{argument_list} which spans
from the line 766, column 28 to line 766, column 66. The spans of the
named captures (\e{@pre}, \e{@wrong_arg}, \e{@post}) are also given. The feature
\e{@correct_arg} required by the fix is provided as a \e{field_access} node
with the specific children.

\subsection{Specifying examples} \label{examples}
An \e{example} specification represents the actual code change applied
to fix a bug in a code base.

The code change is represented by two versions of the code: the buggy
``before'' version and the fixed ``after'' version.

If the code base is stored using Git, Apache Subversion or another compatible
version control system, then the \e{example} specification can specify the
``before'' and ``after'' versions as commit identifiers.

\begin{lstlisting}
example CLOSURE_14
repository
  https://github.com/google/closure-compiler
before
  b7c2861bf45b358b26ebc5ee1be9b6ce96bec78a
after
  4b15b25f400335b6e2820cb690430324748372f9
language
  JAVA
hunk
diff --git a/src/com/google/javascript/jscomp/ControlFlowAnalysis.java b/src/com/google/javascript/jscomp/ControlFlowAnalysis.java
index 5c6927f9c08..980deff1df6 100644
--- a/src/com/google/javascript/jscomp/ControlFlowAnalysis.java
+++ b/src/com/google/javascript/jscomp/ControlFlowAnalysis.java
@@ -764,7 +764,7 @@ private static Node computeFollowNode(
        } else if (parent.getLastChild() == node){
          if (cfa != null) {
            for (Node finallyNode : cfa.finallyMap.get(parent)) {
-              cfa.createEdge(fromNode, Branch.UNCOND, finallyNode);
+              cfa.createEdge(fromNode, Branch.ON_EX, finallyNode);
            }
          }
          return computeFollowNode(fromNode, parent, cfa);
end
\end{lstlisting}

\subsection {Specifying fixes and applications together with a bug} \label{together}
In the preceding examples the bug, fix and application specifications are separate. There are two reasons for enabling this possibility: 
\begin{itemize}
\item Several fixes and applications may be available for a single bug.
\item For a bug specified in one place, bug fixes may be proposed in many different places, by authors who do not have the ability or permission to modify the original bug specification; the same observation applies to applications.
\end{itemize}

In some cases, however, it may be convenient when describing a bug to specify one or more fixes, as well as one or more applications, directly with it. The syntax support these possibilities by allowing one ore more \e{fix} clauses, and one or more \e{application} clause, after the \e{where} clause of a \e{bug} specification. Here is a case involving a bug-with-fix specification, equivalent to the separate specifications given above, with an \e{application} clause also added:

\begin{lstlisting}
bug WRONG_ARGUMENT_IN_CALL_2 
parameter
  @correct_arg: EXPRESSION -- This is a feature
                               -- of the fix
where
  (ARGUMENT_LIST
    (_)* @pre
    (_) @wrong_arg
    (_)* @post)
fix CORRECT_ARGUMENT_IN_CALL_2 
then
  (ARGUMENT_LIST
    @pre
    @correct_arg
    @post)
application
  @correct_arg = (identifier "a")
end
\end{lstlisting}

To specify more than one fix, it suffices to add more \textbf{fix} clauses, each with its name (such as \e{CORRECT\_ARGUMENT\_IN\_CALL\_2} here). In the same way, any number of \e{example} clauses may be present. While such a specification always starts with a single \e{bug} clause, the order of \e{fix} and \e{example} clauses, if any, is arbitrary, although the recommended order is: bug, fixes if any, examples if any.

\section{Constructs and languages} \label{bugfix2}

The bug specifications as illustrated above, and through them the example and
fix specifications, refer to language constructs. The second part of \toolname includes a way to describe the corresponding constructs, at two levels:

\begin{itemize}
    \item In a language-independent way: for example you may specify the concept of loop, in one of its variants such as the ``while loop'' available in some programming language. Such a \textbf{construct} specification indicates the components of a loop, for example the initialization part \e{i}, condition \e{c} and body \e{b}. Such a description does not prescribe any particular programming language.
    \item In language-specific way: for example, a while loop in the C language is of the form
            for \{\textit{i} ; \textit{c} ; {\textit{b}}\}
        where the elements in italics correspond to the components of a particular loop.
\end{itemize}

The following two subsections describe these mechanisms.
 
 \subsection {Specifying constructs} \label{constructs}

A construct specification simply indicates how a particular programming construct is made up of elements. 

Terminology: a ``construct'' is a language concept, such as \e{WHILE_LOOP} below. A program in the language is made up of ``specimens'' (instances) of those constructs. For example, a particular while loop is a specimen of the construct \e{WHILE_LOOP}.

\subsubsection{Describing a construct with required components} \label{components}

The following example describes ``while'' loops:

\begin{lstlisting}[captionpos=b, basicstyle=\fontsize{0.27cm}{0.27cm}]  
construct WHILE_LOOP 
kind
  INSTRUCTION
feature
  body: INSTRUCTION
  condition: EXPRESSION
  initial: INSTRUCTION 
end
\end{lstlisting}

\noindent The basic idea is simply to express a construct as consisting of a number of named components, describing the abstract syntax. Here a \e{WHILE_LOOP} includes a body, a condition and an initialization part, of respective types \e{INSTRUCTION}, \e{EXPRESSION} and \e{INSTRUCTION} again. The order of their listing is irrelevant since the specification only addresses abstract syntax; order will only become significant in the concrete syntax specification for a particular language (\e{language} specifications, section \ref{language}); that will be the place to specify that, in a particular programming language, a loop has the form \e{from initial while condition do body end}.

\subsubsection{Alternation constructs} \label{alternation}

The \textbf{kind} clause lists a construct, \e{INSTRUCTION}. Its semantics is that the construct being defined here, \e{WHILE_LOOP}, is a particular case of another construct \e{INSTRUCTION}; the idea is similar to inheritance in object-oriented programming (where a compiler could have a class \e{WHILE_LOOP} inheriting from an abstract class \e{INSTRUCTION}).

Such constructs are not permitted to have their own \e{construct} specifications but are declared contextually by appearing in the \e{kind} clause of at least one \e{construct}.

This convention is important for the flexibility of the language description
since it enables each language to pick its own variants. It applies to
``alternation'' constructs such as \e{INSTRUCTION} which in traditional BNF
would be defined through the specification of a set of alternatives, as in
\e{INSTRUCTION = ASSIGNMENT \| WHILE_LOOP \| CONDITIONAL \| ...}. Such an
approach would, however, defeat the \toolname purpose of  specifying programming mechanisms in a language-independent way, and then, separately, their combination and rendition in a particular programming language:

\begin{itemize}

    \item 
    While all imperative programming languages have some kind of instruction construct with a number of variants, they differ in their exact choice of variants. For example, while ``for loops'' and ``while-do'' loops appear in both Java and C, Java does not have the C ``do-while  ``while loops''.

    \item 
    If the \toolname specify a construct such as \e{INSTRUCTION} by listing its alternatives, that construction would need many different variants, such as \e{C_INSTRUCTION} and \e{JAVA_INSTRUCTION}. But then it would become impossible to define other constructs such as \e{WHILE_LOOP} since some of their components such as \e{body} must refer to \e{INSTRUCTION}.
    
    \item 
    The \toolname solution is instead to consider such alternation constructs as defined simply by their appearance in at least one \e{kind}  clause for a \e{construct} specification.

    \item 
    For a particular programming language, the alternation is defined as a choice between all such constructs included in the \e{language} specification (section \ref{language}). 
    
\end{itemize}

 \noindent With this convention, an alternation construct such as \e{INSTRUCTION} becomes a construct without an explicit specification, simply by appearing in \e{kind} clauses for some other constructs. 

\subsubsection{Optional and list components} \label{option}
The following example completes the illustration of how to specify constructs in \toolname, by introducing  two other facilities: optional and list components. 

\begin{lstlisting}[captionpos=b, basicstyle=\fontsize{0.27cm}{0.27cm}]  
construct GENERAL_CONDITIONAL 
kind
    -- A conditional instruction 
    --   with a "then" part,
    --   0 or more of "else if"
    --   and an optional "else" part.
  INSTRUCTION
feature
  condition: EXPRESSION
  then_part: CONDITION_INSTRUCTION
  else_if*: CONDITION_INSTRUCTION
  else_part?: INSTRUCTION
end

construct CONDITION_INSTRUCTION 
kind
    -- A <condition, instruction> pair.
  INSTRUCTION
feature
  condition: EXPRESSION
  instruction: INSTRUCTION
end
\end{lstlisting}

The question mark ``\e{?}'' after a component name, as in \e{else_part?},
indicates that the component is optional: a specimen of the construct may,
or not, include a specimen of that component.

An asterisk after a component name denotes a sequence (list):
here the \e{else_if} component consists of zero or more specimens of
\e{CONDITION_INSTRUCTION}.
Since zero specimen is a possibility, the component may be absent.
To specify one or more, use ``+'' instead of ``*''.

The first construct definition above defines a ``general conditional'' as consisting of: an expression, called \e{condition}, required; an expression, the \e{then_part}, also required; zero or more ``else if'' parts, each made of a condition and an instruction (as specified by the auxiliary construct \e{CONDITION_INSTRUCTION}); and an optional ``else part''.

\subsection {Specifying languages with existing grammars} \label{language}

To define a language in a very formal way, exsting Tree-sitter grammars can be
used. The approach for defining a language this way is similar to defining
bug-fix pair in \toolname.

Each language entity is defined using 4 different kinds: a \textit{source} (the Tree-sitter query in the
concrete language grammar); a \textit{construct} (the pattern for constructing the \toolname \e{construct}); a \textit{construct\_name}; a \textit{language}.

For example, the following code extract shows how \e{ARRAY_ACCESS}
construct can be extracted from Java and Python programs:
\begin{lstlisting}
language 
    JAVA 
construct_name 
    ARRAY_ACCESS
source
  ASSIGNMENT_INSTRUCTION:
      (ARRAY_ACCESS
        array: (_) @array
        index: (_) @index)
construct
    (ARRAY_ACCESS
        array: @array
        index: @index)
end

language 
    PYTHON
construct_name 
    ARRAY_ACCESS
source
      (SUBSCRIPT
        value: (_) @array
        subscript: (_) @index)
construct
    (ARRAY_ACCESS
        array: @array
        index: @index)
end
\end{lstlisting}

The nodes captured in the query pattern and used in the construction pattern
are translated recursively.

\section{\toolname website and API} \label{API} 
The \toolname specifications are available as a website at \sitelink
which has both the human-readable pages and the machine-readable JSON API.
The website's data takes advantage of the already cited earlier  sketch,  ~\cite{aprlisbon}.

The website is built using the static website generator
Jekyll~\cite{preston2018jekyll} which generates both the HTML code for the webpages
and the JSON files for the API. The input for the generator is a collection
of Jekyll data files which contain specifications in the \toolname language
format. This allows the website to be used for exchanging information
related to different projects.

Both the API and the webpages provide the same
data, so the following description focuses on the API view.

The website has a Jekyll collection for each kind of \toolname elements. Each
collection has the index page which lists all elements of this collection. For
example, the \texttt{/bugs} page lists all \e{bug} elements present in the
database. This page and all other pages have the JSON API counterpart which
is located at the same path but with the \texttt{.json} extension appended
(in this case it is \texttt{/bugs.json}). Same for \texttt{/fixes},
\texttt{/applications}, \texttt{/examples}, \texttt{/constructs}, \texttt{/kinds}
and \texttt{/languages} for language elements (the full list of collections
is visible on the website).

Each element has an ID. Index JSON routes respond with a JSON object
which has element IDs as keys and the main fields of the elements as values.

The extended information of each element is available on the element's page
in the collection. The path to the element's page is constructed by
appending the element's ID as a path segment. For example, one of the bugs in
the database has an ID \texttt{null\_check\_missing\_1}, so the page of the
bug is located at \texttt{/bugs/null\_check\_missing\_1} and the JSON API
is located at \texttt{/bugs/null\_check\_missing\_1.json}.

The fields of the JSONs reflect the clauses of the \toolname specifications of
the respective element kind. The clauses containing the sexp-based patterns are
available both as a string and as JSON objects to simplify the creation
of tools using \toolname. For example, the JSON of the bug
\e{NULL_CHECK_MISSING_1} has the \e{"where"} field for the parsed \e{where}
pattern and the \e{"where_raw"} field for the raw string with the S-expressions:

\begin{lstlisting}
{"id": "wrong_return_statement_1",
"where": {
  "type":"return_statement",
  "children":[{
    "field":"return_value",
    "node":{"type":"true"},
    "name":"@wrong_value"}]},
"where_raw":
  "(return_statement return_value: (true) @wrong_value)",
"fixes": ["correct_return_statement_1"]}
\end{lstlisting}

The webpage versions of the pages include hyperlinks in the pattern fields
for the ease of navigation.




\section{Related work} \label{related_work}
The most important precursor to the design presented here is the important work
pursued by several groups, over the past decades, to provide repositories of
bugs and bug fixes; this concerted effort is a prerequisite to any attempt at
enabling teams working on Automatic Program Repair to assess and compare their
methods. Some of the most significant such repositories are the following \cite{campos2017common, soto2016deeper,livshits2005dynamine,williams2005automatic,sun2017empirical,duraes2006emulation,catolino2019not,lin2017quixbugs,sobreira2018dissection}.

Another precursor is the work on code analysis tools. Boa \cite{dyer2015boa} is a
domain-specific language and infrastructure for analysing software repositories.
Boa has a common AST structure for Java, Python and Kotlin programs present
in the Boa datasets. To analyze the AST of programs in other languages programs
must be translated into Boa AST, this was done in \cite{biswas2019boa} for
Python. By default, Boa AST is not extendible. A separate project Boidae
\cite{sigurdson2024boidae} allows extensions but they must be done directly
in the Java source code of the system.

Research in multilingual code analysis can use several distinct AST structures
such as \cite{nielebock2019programmers} which combines results
of several parsers and \cite{rix2024semantic,clem2021static}
in which the Semantic \cite{rix2024semantic} does code analysis on different
ASTs produced by different tree-sitter parsers.
Another approach is to to analysis on a pre-existing common AST as in
\cite{keshk2023method} which uses the Boa AST for Java, Python and Kotlin.

A short paper \cite{aprlisbon} was to our knowledge the first to propose a possible framework for describing bugs and fixes. It is, however, very short and more of a position paper, whose results are not directly usable; in addition, unlike the present work, it does not provide any directly usable resources.

ASTs are produced by parsers according to specified grammars. Parser generators
such as ANTLR \cite{parr1995antlr} and others have been used both in language compilers
and in research \cite{latif2023comparison}. Tree-sitter
\cite{tree-sitter} is a parser generator which has language grammars
available for many languages. Tree-sitter is used in code editors and code
analysis research \cite{le2022hyperast,nielebock2019programmers,clem2021static}.
Another feature of tree-sitter is its query language. The query language allows
specifying subtree patterns which are found in the parsed AST.

\section{Assessment, limitations, conclusions and future work}  \label{future_plans}

Departing from the usual practice of describing specific techniques in testing and Automatic Program Repair, the present work positions itself at what could be called a ``meta'' level: it abstracts from individual research projects to provide instead a set of concepts, tools and resources for everyone working in the field.

The authors' own evaluation has consisted of considering all the bug and fix repositories which we could get hold off, more specifically all the repositories covering \textit{coding} bugs and fixes. and painstakingly ensuring that the notation and tools could cover all their contents simply and effectively.

By that measure, the effort is successful, since we have \toolname descriptions for all these examples from the literature and widely used repositories. Full validation can only come from widespread use by researchers and practitioners working in complete independence from the authors.

While it would be desirable to cover more sophisticated kinds of bugs, such as design bugs, the restriction to the coding case seems justified think such further advances will only possible if we have a fully satisfactory solution to this application, covering the bugs that are best understood and on which APR techniques have produced their most significant results so far.

The contributions to the advancement of Automatic Program Repair described in this article include:

\begin{itemize}
    \item A standard language for describing \textit{bugs and their fixes}.
    \item Supporting it, a standard language for describing \textit{programming constructs} from virtually any programming language.
    \item For these languages, use of \textit{widely accepted techniques}, tree-sitter parsing and querying access to the information in a standard way (unlike the approach of \cite{aprlisbon} which only used new notations).
     \item Along with all these (human-oriented) notations, \textit{API (programming interface) versions} making the functionality accessible to any production or research tool.
    \item The associated \textit{database mechanisms}.
    \item An \textit{actual database}, publicly  avaiable, providing a repository of important bugs and fixes collected from a wide variety of sources.
    \item A working \textit{website} providing access to all this information, both in human-readable HTML and machine-readable JSON.
    \item For all these applications, an \textit{open, modular approach} making it possible to add new constructs, new languages, new kinds of bugs and fixes and new data sets in a flexible way.

\end{itemize}

The present work should be viewed as a \textbf{community-service contribution}: it is meant to help everyone interested in bugs and their automatic repair, particularly researchers in the field, to achieve new advances by relieving them of common and repetitive tasks,  facilitating the assessment of new techniques, and making it possible to compare the effectiveness of competing or complementary approaches by providing a common reference. 

Unlike previous attempts, which were just sketches (albeit in the right direction), the proposal presented here, while still incomplete, provides enough information (in particular, the notations and API) and resources (in particular, the repositories) to enable actual use by bug-analysis and Automatic Program Repair projects.

We have made every effort to make these tools and resources attractive and useful to the community, and will be happy to adapt them in response to user feedback. The most significant evaluation will come from such use.

Just as understanding bugs is fundamental to software engineering, progress in Automatic Program Repair is fundamental to the progress of practical software development. We strongly hope that the community will find \toolname a valuable resource and that it will provide a new impetus to further advances in the field.   

\bibliographystyle{IEEEtran}
\bibliography{reference}
\end{document}